# Reconfigurable, Multifunctional Origami Electronic Membranes for Mechanical and Environmental Sensing

Yao Yao[1+], Guanghui Li[2+], Xin Ning[1*]

1. Department of Aerospace Engineering, University of Illinois at Urbana-Champaign, Urbana, 61801

2. Department of Aerospace Engineering, Pennsylvania State University, University Park, 16801

* Corresponding author: xning@illinois.edu

+ Equal contribution

## Abstract

This work introduces a concept of origami electronic membranes that leverages the design and fabrication of flexible electronics and the mechanical behavior of engineering origami to achieve unique multifunctional, shape-reconfigurable, and adaptive membranes for mechanical and environmental sensing in benign and harsh conditions. This paper presents the materials, design, and fabrication methods for realizing six origami electronic membranes capable of reconfiguring planar or three-dimensional shapes based on the modified flasher, Kresling, Miura-ori, circular, letter, and Tachi-Miura origami patterns. These origami-based, thin-film flexible electronics can obtain both expansion and folding of their shapes, as well as transformation between different geometries. The origami electronic membranes can achieve mechanical and environmental sensing functions such as measuring motions, mechanical strains, temperatures, UV light, and humidity. The results reported here demonstrate the promise of combining engineering origami with flexible electronics to advance the state-of-the-art in multifunctional foldable and deployable electronics and systems.

## 1. Introduction

We report a concept of origami electronic membranes (OEMs) based on merging the design and fabrication of flexible electronics and the mechanical behavior of engineering origami to achieve unique multifunctional, shape-reconfigurable, and adaptive membranes for mechanical and environmental sensing in benign and harsh conditions. Flexible or soft electronics can adapt to high-strain deformations and conform to three-dimensional (3D) complex geometries,[1-3] offering wide applications in wearable electronics,[3, 4] biomedical devices,[5] integrated circuits,[6] energy storage,[7] and many others. Recent efforts have made progress in increasing the shape complexity for improved functions by using multilayer devices,[1] developing 3D curvy electronics via conformal additive stamp printing technology,[2] and assembling 3D electronics by compressive buckling.[8] Despite these advances and the advantage of allowing stretching, an intrinsic challenge with flexible electronics is that they induce mechanical strains and stresses on the functional devices during stretching and may require high forces at large strains. This issue limits the applications of flexible electronics in scenarios that require shape expansion beyond 100% since the typical allowable strains of flexible electronics are largely below this threshold. It is also difficult to compress flexible electronics into a smaller size because they are typically not designed for compression which may create damage and wrinkles. There is a need for flexible electronics that can shrink dimensions in a repeatable and controlled manner.

Here, we leverage the principles and designs of engineering origami to address these issues with flexible electronics. Origami is an ancient art of paper folding that folds two-dimensional (2D) sheets along the prescribed creases to create 3D objects. Extensive work has explored the engineering applications of origami in aerospace,[9] architecture,[10] robotics,[11] actuators,[12] metamaterials,[13] medical stents,[14] and many others. For example, Lin et. al developed origami paper photodetector arrays that achieved 1000% deformation.[15] However, print paper is not a durable engineering material. Origami lithium-ion batteries were created by combining a Miura-ori structure with energy storage cells, achieving significant linear and areal deformability.[7] Origami solar panels have been proposed as new space solar wings due to their high packaging efficiency.[9] However, only a limited group of origami designs have been studied with a heavy focus on the Miura-ori folding scheme. The potential of engineering origami in advancing flexible electronics is yet to be fully explored. A unique feature of origami designs is that they utilize the motions of their facets around the creases to achieve significant and complex shape morphing, expansion, or compression, allowing zero or minimal strains or stresses induced on the facets. Therefore, origami designs are promising solutions to achieve large shape expansion or compression without causing the issues of traditional flexible electronics. In this work, we combine a large group of engineering origami designs with state-of-the-art microfabrication techniques and use advanced engineering materials to obtain origami-based, multifunctional flexible electronics capable of expanding, shrinking, or transitioning between complex shapes.

## 2. Results

**2.1. The concept of origami electronic membranes**

Origami can expand into a large structure from a small volume with zero or minimal strains induced in its facets, offering a promising approach to applications that require shape reconfiguration or deployment. We leverage this advantage of origami and combine the structural design with flexible electronics to realize shape-reconfigurable multifunctional devices. Here, a multifunctional and shape-reconfigurable origami structure based on a modified flasher pattern[16] serves as an example to present the design, functions, and fabrication of origami electronic membranes, referred to as OEMs (Figure 1). A modified flasher origami pattern utilizes radially distributed hill and valley folding lines, respectively shown as the solid and dash lines in Figure 1a, to obtain radially wrappable and deployable structures. Supporting Figure S1 a presents the detailed folding crease pattern of the modified flasher OEM. Figure 1a presents this OEM from a deployed configuration (state 1) to a fully wrapped shape (state 4).

An OEM is an integrated structural-electronic system consisting of four basic layers and an optional encapsulation layer (Figure 1b). A 20-µm-thick polyimide (PI) layer serves as the thin-film structural supporting material to form and integrate the circuits and functional electronics. We choose PI as the structural substrate because of its durability and wide applications in engineering applications[17] and even in some extreme environments such as outer space[18]. A circuitry or electrode layer with an encapsulating film is patterned on the PI substrate to power and connect the electronic components. The electronic components can be off-the-shelf commercial microchips and/or customized, micro/nano-fabricated thin-film devices. Depending on the specific application, an additional encapsulation can be added to protect the electronic components. In this work, we have demonstrated the capabilities of environment sensing for light (ultraviolet, UV), humidity, and temperature, as well as the monitoring of mechanical strain and motions (Figure 1c). Table S1 summarizes all the sensing strategies and respective sensing materials. As representative examples, we use the OEM based on modified flasher origami to show the functions of motion sensing, temperature measurements, and UV responses.

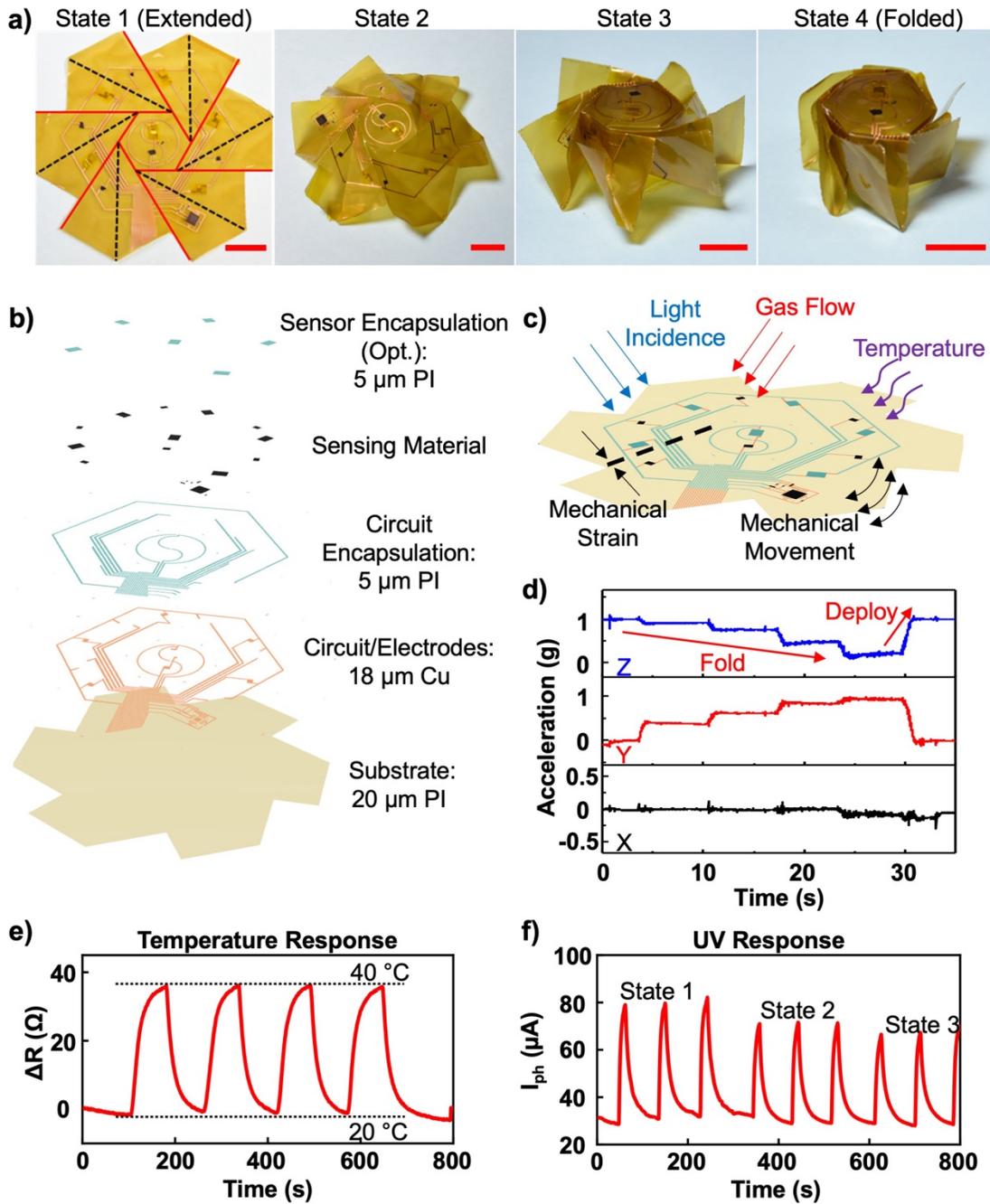

**Figure 1. Integration of flexible electronics with origami membranes for reconfigurable origami electronics.** a) Optical images of an origami electronic membrane based on a modified flasher pattern in extended and fully folded configurations as an example of radially reconfigurable electronics. Scale bars: 15 mm. b) Exploded schematic of the materials and devices of the origami electronic membrane. c) Possible sensing modalities of the origami electronic membranes. d) Measurements of the motion of origami membrane during folding and deploying. e-f) The responses of the temperature and UV sensors in various folding configurations.

A surface-mount, off-the-shelf commercial accelerometer (ADXL335) serves as the sensor for monitoring the motion of the origami facet during folding and deployment. Figure S2 includes the location of the accelerometer and its orientations. The z-axis of the accelerometer is perpendicular to its surface and the origami facet, and the x- and y-axes are parallel with the facet. The z-axis is along the gravity when the origami is fully extended, resulting in a measurement of 1 g in state 1. As the origami folds into state 4, the z-axis tilts toward the horizontal plane. Therefore, the z-axis reading reduces to 0g in the fully folded shape. In contrast, the y-axis tilts from horizontal to vertical during this process (0 to ~1g). The origami OEM is then fully wrapped back into state 1, and the signals in all three directions recover to the initial conditions.

Thin-film temperature and UV sensors serve as the devices for monitoring the external environmental conditions of the OEM. The temperature sensors are serpentine, 200 nm thick copper (Cu) micro-traces that can change the electrical resistance with variable temperatures (Figure S3). The base resistance (i.e., the value at room temperature) can be designed by modifying the Cu trace's thickness, width, and length. Measuring the variations in the electrical resistance provides the means for sensing the external temperature. Figure 1e shows the resistance variation of the temperature sensor on the modified flasher OEM under thermal cycles between 20 $^{\circ}$C and 40 $^{\circ}$C. The UV sensor uses photo-sensitive nanocomposites of zinc oxide (ZnO) and carbon nanotube (CNT) that have been extensively studied in our previous work. The UV light on the sensor can generate current, i.e., photocurrent, which the Cu electrodes collect (Figure S4). The photocurrent increases as the incident UV light increases, serving as the mechanism for sensing the intensity of UV light. Figure 1f includes the photocurrent responses of the UV sensor at different shape states. Although we envision that the UV sensors are mainly used at the fully deployed configuration to provide UV measurements, the photocurrent responses at partially folded configurations are also included. As expected, the photocurrent decreases as the OEM folds. The results in Figure 1e-f showcase the capability of using origami-integrated, deployable sensors to monitor external environmental conditions such as temperature and UV intensity. Section 3 discusses the details of the materials and fabrication methods of the temperature and UV sensors.

## 2.2. Diverse designs of origami electronic membranes

We extend the concept of OEM to achieve a diverse set of origami designs to showcase its broad applicability in structures with different complexities, including six origami designs categorized into two groups. The first group includes OEMs that can transform between a two-dimensional (2D) shape and a three-dimensional (3D) geometry such as the modified flasher OEM. Another 2D-3D reconfigurable OEM based on the Kresling origami[19] is also realized (Figure 2a). The Kresling OEM has a pseudo-cylindrical shape that can be axially compressed into a flat shape. The detailed application enabled by Kresling origami's unique shape-reconfiguration capabilities is discussed in Section 2.3.

The second group of OEMs includes origami designs that can transition between a 2D shape and another different 2D geometry (i.e., 2D-2D transformation, Figure 2b). A Miura-ori OEM leverages the classic Miura-ori folding pattern[20] and integrates sensor arrays for environmental and mechanical sensing. This Miura-ori OEM uses coupled folding or deployment in biaxial directions to synchronously retract or deploy sensor arrays. It can transition between a fully extended planar surface and another flat shape in a more compact configuration. Moreover, we use two simple origami patterns – a circle origami[21] and a letter folding origami[22] to obtain OEM to show additional designs that can transform from a smaller 2D configuration to a larger, expanded surface. The last OEM in this work is based on the Tachi-Miura pattern[23] that transitions between vertical and horizontal planes. Figure S1 presents all the crease patterns of these OEMs. Subsequent sections will discuss the applications and functions of these OEMs.

| | Origami Pattern | Fully Deployed | Intermediate State | Fully Folded |
|---|---|---|---|---|
| a) 2D ↔ 3D | Modified Flasher | 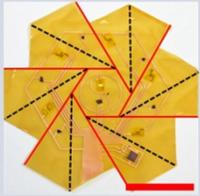 | 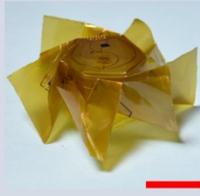 | 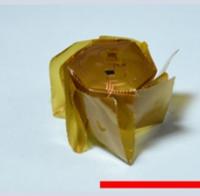 |
| | Kresling | 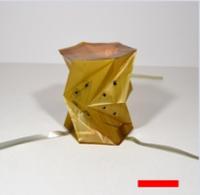 | 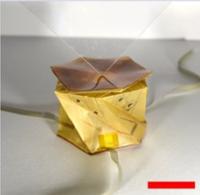 | 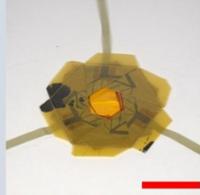 |
| b) 2D ↔ 2D | Miura-Ori | 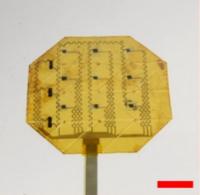 | 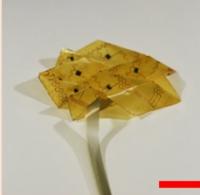 | 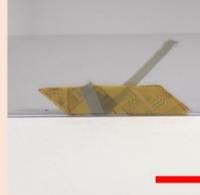 |
| | Circle | 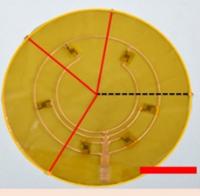 | 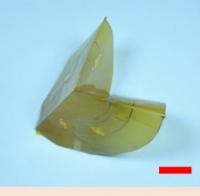 | 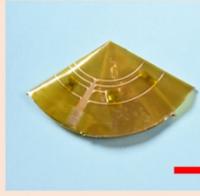 |
| | Letter | 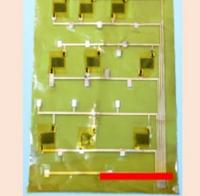 | 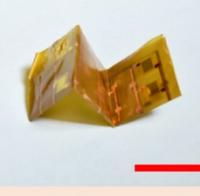 | 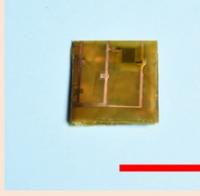 |
| | Tachi-Miura | 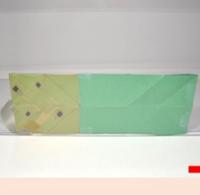 | 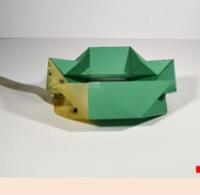 | 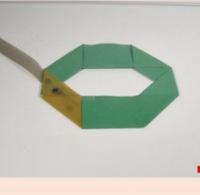 |

Figure 2. Summary of origami electronic membranes developed in this work, including designs that can transform between a) 2D and 3D geometries and b) different 2D shapes. Scale bars: 30 mm

## 2.3. Function-switchable origami electronic membranes enabled by shape reconfiguration

We use the unique shape-reconfigurable capability of the Kresling origami pattern to realize switchable functions based on changing the shapes to adapt to different needs for sensing modalities. A Kresling origami consists of axial tessellations of pseudo-cylindrical units that are comprised of repetitive triangles. The unit's twisting motion around its axis and the axial compression are coupled; therefore, twisting a Kresling unit can axially collapse or extend it, and axial compression or extension will lead to its rotational twisting. The twisting directions of the units in a Kresling tessellation can be arranged in identical, opposed, or mixed directions to achieve tailorable rotation behavior when the Kresling structure is axially deformed.

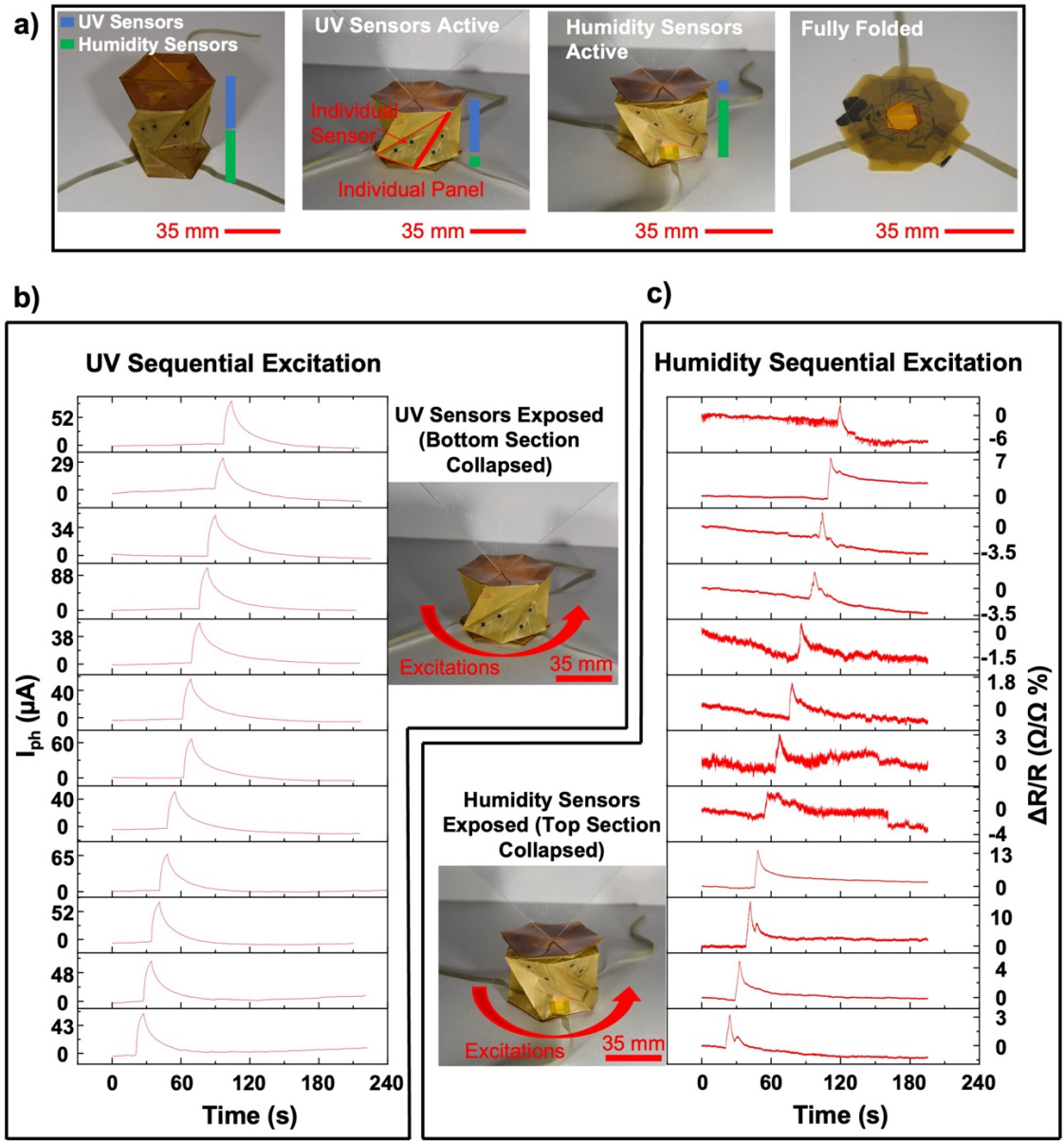

**Figure 3. Origami electronic membranes as functionally switchable and axially collapsible sensing platforms.** a) Optical images of functionally switchable origami electronic membrane based on Kresling pattern at various configurations. b) Responses of each UV sensor under sequential excitation. c) Responses of each humidity sensor under sequential excitation.

The Kresling OEM in this work consists of two units arranged in a mirror-symmetric configuration about the middle, resulting in opposed twisting directions when axially compressed (Figure 3). Each unit has 12 identical triangles that are folded into a pseudo-cylindrical configuration (Figure S1 b). When the Kresling OEM is twisted in the clockwise direction, the bottom unit will be fully folded and the unit on the top will remain extended. Likewise, twisting the Kresling OEM counterclockwise will fold the top unit and keep the bottom unit extended (Figure 3a). Applying axial compression while allowing the units to rotate can fully fold the entire structure (Figure 3a). Supporting video 1 shows the shape reconfiguration process. This capability of shape reconfiguration enables the switching of sensing functions of the Kresling OEM.

The top unit of the Kresling OEM has an array of 12 UV sensors, and the bottom unit integrates 12 humidity sensors. The materials and fabrication methods of the UV sensors are the same as the ones in the modified flasher OEM. The humidity sensors use reduced graphene oxides (rGO) as the active sensing material. As the humidity increases, the resistance of rGO also increases. Figure S5 includes the detailed design of the rGO humidity sensors and Section 3.4 presents the fabrication methods. When the Kresling OEM is twisted in a clockwise direction, the top unit remains at the extended configuration to expose the UV sensors while the humidity sensors in the bottom unit are fully stowed. Therefore, the OEM is now programmed into a UV-sensing device. We sequentially illuminate the UV sensors around the OEM with a UV LED. They generate clear photocurrent from the incident UV light, and their photo-responses are in Figure 3b. These results indicate that this OEM may be used as a UV sensing device that can measure UV irradiation from 360° directions. Twisting the Kresling OEM in a counterclockwise direction will expose the humidity sensors at the bottom and hide the UV sensors on the top, switching the OEM into a humidity-sensing device. We sequentially blow humid air to the humidity sensors around the structure, and their variations in electrical responses are in Figure 3C. These results indicate that this OEM may be used as a humid-sensing device to measure humidity in the air from 360° directions. Both the UV and humidity sensors can be used concurrently if the top and bottom units are fully extended by stretching the Kresling OEM. The entire structure can be fully compressed into a flat shape for small stowage volume when it is not used.

**2.4. Biaxially deployable origami electronic membranes for UV sensing in harsh temperatures**

An OEM based on Miura-ori serves as a sensing platform that can extend into a large area from a small, stowed volume (Figures 2b and 4). Supporting video 2 shows the fully folded Miura-ori passing through a narrow space and then flattened into a 2D surface. This OEM has 3 strain sensors on three hill creases and 9 UV sensors on various facets. The strain sensors use nanocomposites of rGO and carbon nanotubes (rGO/CNT) as the sensing material with a high gauge factor of ~50[24]. Their electrical resistance can increase when rGO/CNT nanocomposite is stretched, enabling the sensing of strains by measuring the resistance. Figure S6 shows the design of the strain sensors, and Section 3.4 discusses the fabrication methods. When the Miura-ori OEM transitions from the initial state (Figure 4b, top) to the fully folded configuration (Figure 4b, bottom), the strain sensors' resistances increase to the maximum (Figure 4c). The resistances recover after the Miura-ori OEM is released back to its deployed state. The stain sensors may be used to monitor the folding status of the Miura-ori OEM.

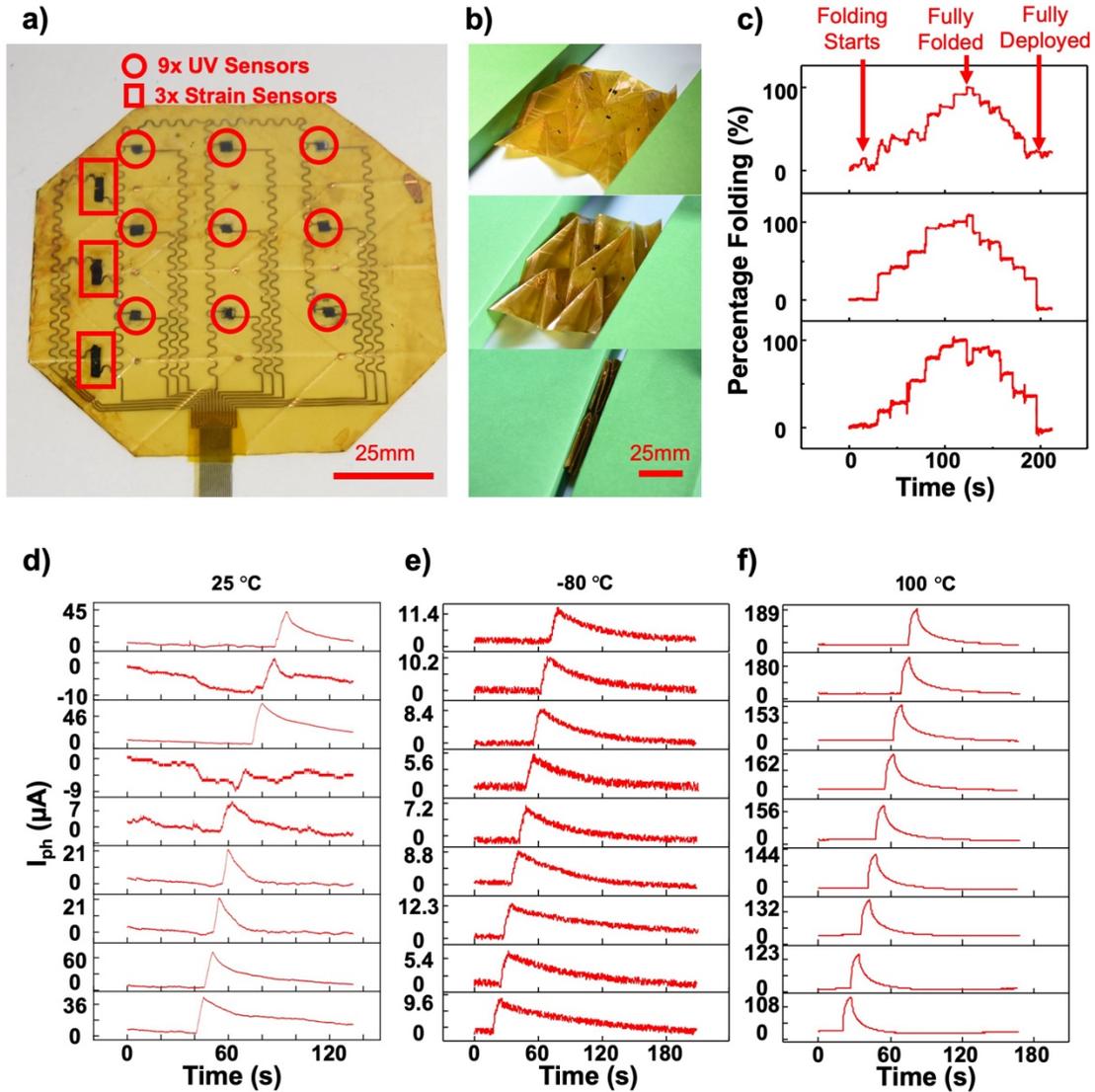

**Figure 4. Origami electronic membranes as biaxially reconfigurable UV sensing platforms in austere environments.** a) Origami electronic membrane based on Miura-ori pattern. b) Folding of the Miura-ori origami electronic membrane. c) Strain response during the folding and deployment process. d) Photocurrent responses of the thin-film UV sensors at room temperature. e-f) Photocurrent responses of the thin-film UV sensors at -80 °C and 100 °C, respectively.

The UV sensors utilize the same materials and fabrication methods as discussed in the previous sections. We sequentially illuminate the UV sensors, which generate photocurrent upon the UV irradiation. Figure 4d presents the photocurrent responses of the 9 UV sensors under sequential excitation at room temperature. To demonstrate the sensing function at extreme temperatures, we put the Miura-ori OEM in a cold condition at -80 °C generated by a thermal chamber and a hot environment at 100 °C on a hot plate with UV light sequentially illuminated on the sensors. Figure 4e-f presents the photocurrent at low and high temperatures, respectively. The results show clear photoresponses under UV excitation, suggesting that the Miura-ori OEM may be used as a deployable UV sensing surface in extreme temperatures.

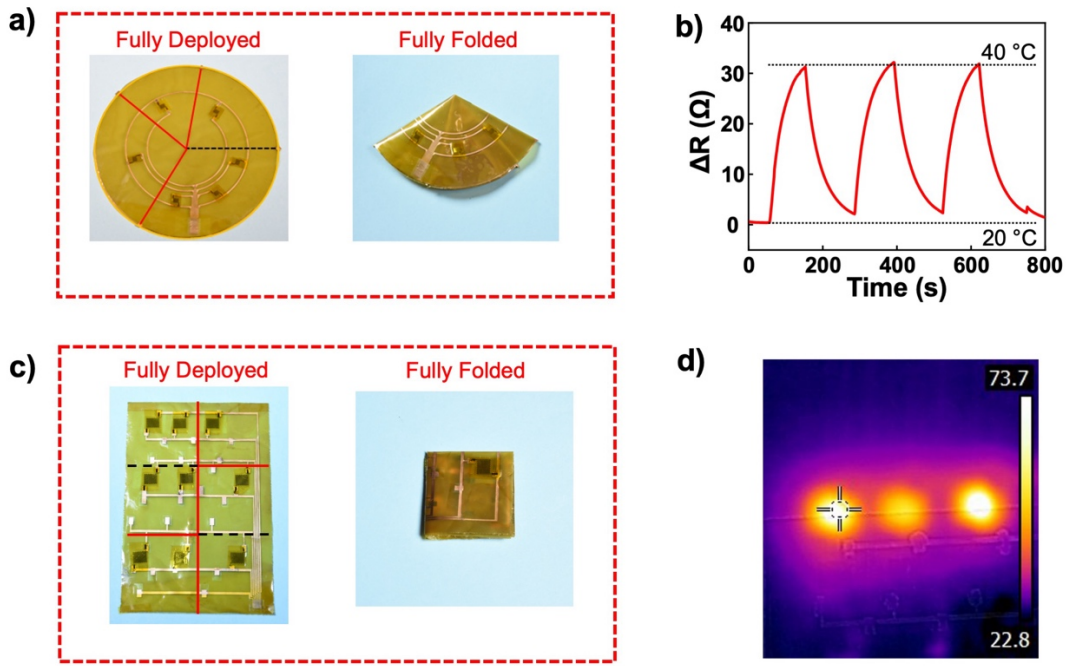

**Figure 5. Origami electronic membranes as 2D expandable temperature sensing and control platforms.** a) Optical images of a circular temperature sensing membrane. b) The variation of resistance of a temperature sensor at variable temperatures. c) Optical images of a letter-folding origami electronic membrane with micro-heaters. d) An infrared image of the microheaters in expanded configurations.

**2.5. Deployable origami electronic membranes for temperature sensing and control**

We further demonstrate the application of OEMs as deployable thermal sensing and manipulation devices. Two origami designs based on a generic circular origami pattern and a letter folding pattern serve as the structural designs for the OEMs (Figure 5). The circular origami pattern has three hill creases and one valley crease meeting at its center, following the 3-to-1 rule of rigid origami folding.[21, 22] The circular OEM includes six thin-film thermo-resistive temperature sensors. Figure 5b shows the variation of the resistance of a temperature sensor at cyclic temperatures between 20 °C and 40 °C. Figure 5c shows a letter folding OEM in deployed and folded configurations. The letter folding OEM includes an array of thermo-resistive microheaters that can be heated when powered. Section 3.5 discusses the detailed fabrication methods for the microheaters. Figure 5d is the infrared image of three microheaters when powered on, showing a maximum temperature of 73.7 °C with a 7W energy input. These results showcase that the OEM with integrated thin-film temperature sensors and microheaters may be used as deployable thermal sensing and manipulation platforms.

## 2.6. Origami electronic skins for deployable structures

The OEMs can be used as electronic skins integrated with origami deployable structures to monitor the host structure's motion. Figure 6 presents a Tachi-Miura OEM with two surface-mount accelerometers and two surface-mount gyroscopes. Figure 6a includes the optical images of the Tachi-Miura OEM and its host structure from fully folded to fully deployed configuration. The fully folded shape is a horizontal 2D geometry while the fully deployed state is a flat vertical plane. Figure 6a also presents the circuit design of the Tachi-Miura OEM. Section 3.2 includes the details of the surface-mount accelerometers and gyroscopes and their integration with the origami substrates.

Two accelerometers are integrated on two separate panels the OEM as shown in Figure 6a. During the folding and deployment of the structure, the accelerometers are used to track the relative orientations of the panels of the OEM to the gravity direction. Figure 6b shows the data collected by the accelerometers. At the fully folded state, both accelerometers face downward along the gravity showing a -1 g acceleration in both z-axes, whereas the x- and y-axis readings remain 0 g. As the OEM deploys, the z-axes of the accelerometers rotate to a horizontal position, therefore reducing to 0 g. The x- and y-axes, which are 45 degrees away from gravity, show approximately +/- 0.7 g, depending on the orientation of the chips. The folding after the full deployment resets the axes to their original orientation. During a quasi-static folding or deployment, the linear acceleration measured using the accelerometers can then be used as an indicator of the surface orientation. To demonstrate the capability of measuring surface orientations in the absence of gravity such as in outer space, two gyroscopes are integrated on the OEM on two separate panels (Figure 6a). The gyroscopes measure the rotational velocities of these panels. Integration from the start of the motion yields the cumulated rotational angles compared to the initial orientation (Figure 6c). Before the deployment, the gyroscopes show a 0 °/s in both the x- and y-axes, leading to 0° cumulated rotational angles. After a full cycle of deployment and folding, the cumulated rotational angles return to 0°, indicating that the structure has returned to its initial state.

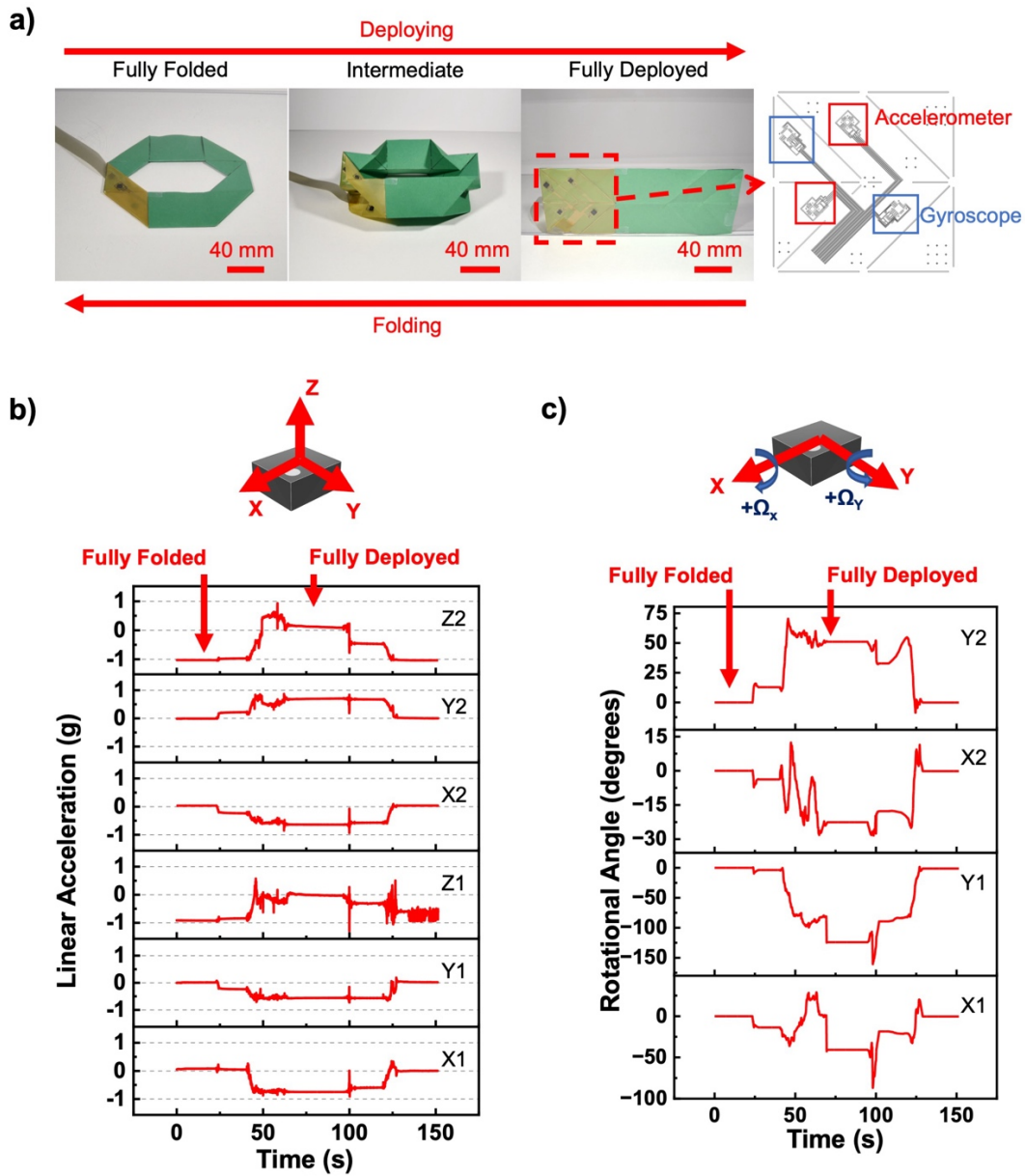

**Figure 6. Origami electronic membranes as electronic skins for monitoring the motions of origami deployable structures.** a) The folding and deployment of the origami structures and the schematic of the attached origami electronic membrane. b-c) Measured surface accelerations and rotations.

## 3. Materials and Fabrication Methods

### 3.1 Fabrication of origami PI substrates with copper circuits

To fabricate an origami substrate with the circuit, a Cu-PI laminate thin film (Pyralux AC182000R) was attached on a Si wafer with a layer of polydimethylsiloxane (PDMS, Sylgard 184, Dow-Corning; 10:1 weight ratio) as the adhesive. Spin-coating photoresist AZ5214 (AZ Electronic Materials) on the Cu-PI film at 3000 rpm for 30 seconds followed by soft baking at 110 °C for 1 minute created a photoresist layer for further microfabrication steps. Exposing the photoresist film under UV (60 mJ/cm2), developing it in AZ 400K 1:4 for 60 seconds, and then baking it at 110 °C for 3 minutes finished the photolithography of photoresist masks for the following etching step. Wet etching in copper etchant type-100 (Transene) created Cu circuits on the PI. Periodically washing the Cu-PI laminate during etching with deionized water accelerated the etching process. The whole etching process took around 15 minutes. After etching, the materials were washed with acetone and isopropyl alcohol (IPA) to remove the photoresist that remained on the Cu. To protect Cu from oxidation in air and balance the thermal stress, a protective layer of PI film was made by spinning PI solution on the Cu circuit at 500 rpm for 30 seconds followed by baking at 140 °C for 4 mins on a hotplate. To expose electrodes to sensing materials, photolithography patterned PI-Cu-PI film by spinning AZ5214 photoresist at 3000 rpm for 30 seconds followed by baking at 110 °C for 1min. Exposing photoresist film under UV (60 mJ/cm2), developing it in AZ 400K 1:4 for 100 seconds to remove photoresist and the uncovered PI film, leading to an open window in the interdigital electrodes. The photoresist was finally washed with acetone and IPA, and the final device was baked at 210 °C in the oven for half an hour.

### 3.2 Accelerometers, gyroscopes, and their integration with the origami substrate

The accelerometer used in this study was purchased from Analog Devices (ADXL335). It is an analog accelerometer capable of measuring linear acceleration of +/- 3.6 g along all three cartesian axes (x-, y-, and z-axis), with a nominal sensitivity of 300 mV/g. The gyroscope used in this study was purchased from STMicroelectronics (LPR503AL). It is an analog gyroscope capable of measuring the rotational velocity of +/- 120 °/s along the in-plane axes (x- and y-axis), with a nominal sensitivity of 8.3 mV/°/s. Output voltages were directly measured from the output pins of the microchips, which were then converted to the actual accelerations and rotational velocities using the respective sensitivities.

The microchips were directly soldered to the copper circuit defined in the previous step. Low-temperature solder paste (TS391LT, Chipquick) was applied to the soldering pad before the placement of the microchips. Heating the sample with a hot plate at 200 °C allowed the melting of the solder paste and aligning the microchips with the copper soldering pads. In case of misalignment, a hot-air rework station was used to apply hot air flow at 220 °C to temporarily melt the solder paste for further alignment of the microchips. Additional electronic components (resistors, capacitors) were soldered using the same method to ensure functionalities.

### 3.3 Fabrication of UV sensors

The ZnO/CNT nanocomposite was synthesized by the solution method described in our previous report.[25] Briefly, 10 mg P2-SWNT (Carbon Solution. Inc) was dispersed in 100 ml 1% sodium dodecyl solution by probe sonication for 1 hour to obtain SWNT solution. 0.82 g Zinc Acetate dissolved in 50 ml Methanol was heated by a hot plate at 60 °C until it was fully dissolved. Then, Potassium Hydroxide (0.5g)/ Methanol (25 ml) solution was slowly and continually added into the Zinc Acetate solution to react for 2 h. This process was repeated three times to obtain ZnO nanoparticles. The resulting ZnO nanoparticles were dispersed in 25 ml methanol. 1 ml CNT solution was added to 1ml ZnO nanoparticle solution followed by bath sonication for half an hour to obtain ZnO/CNT solution. Drop-casting the ZnO/CNT solution on the PI substrate with microfabricated Cu (18 μm thick) electrodes formed a uniform ZnO/SWNT thin film at room temperature. The device was heated at 50 °C in an oven overnight to finish the fabrication of UV sensors.

### 3.4 Fabrication of humidity sensors and strain sensors

Graphene oxide (GO) (Carbon Solution. Inc) was reduced by hydroiodic acid to get reduced graphene oxide. A probe sonication dispersed 10 mg rGO powder in 100 ml deionized water for 0.5 h to get GO dispersion, which was then reduced by N2H2 at 90 °C in a water bath for 12 h. The resulting rGO dispersion was washed with a vacuum filtration method using deionized water and then re-dispersed in 100 ml 1% sodium dodecyl solution by probe sonication for 1h. The humidity sensor was prepared by casting rGO dispersion on the interdigital electrodes on the PI substrate, followed by drying at room temperature and then baking in an oven at 50 °C overnight. To prepare rGO/CNT dispersion, 3mg P2-SWNT powder and 9mg GO powder were dispersed in 100 ml deionized water by probe sonication for 0.5 h. The resulting GO/CNT was reduced and dispersed by the same process as the preparation of the rGO dispersion. The strain sensor was fabricated by casting rGO/CNT dispersion on PI substrate with Cu electrodes followed by drying at room temperature.

**3.5 Fabrication of temperature sensors and microheaters**

The fabrication of the Cu-based temperature sensors and microheaters started with the fabricated Cu-PI substrates. E-beam evaporation method was used to deposit Cu (200 nm) film on the substrate followed by the photolithography method to pattern the 200nm Cu to obtain temperature sensors and microheaters. An additional layer of PI can be spun and baked on the temperature sensors and microheaters to protect them against damage.

## 4. Conclusion

We have developed a new concept of origami electronic membranes (OEMs) based on combining origami designs with flexible electronics that can reconfigure shapes to expand or shrink, switch functions by mechanical shape-reconfiguration, and provide multi-modal mechanical and environmental sensing capabilities. We have developed the materials, designs, and fabrication methods to realize six OEMs capable of 3D-2D or 2D-2D shape reconfiguration based on the modified flasher, Kresling, Miura-ori, circular, letter, and Tachi-Miura origami patterns. The OEMs have been integrated with sensing functions including motion (acceleration and rotation), mechanical strains, temperature, UV light, and humidity. In addition, microheaters integrated with OEMs were also realized for manipulating the temperatures. The modified flasher

OEM was developed to obtain a radially foldable and deployable sensing platform to measure the motions, temperatures, and UV irradiation. The Kresling OEM realized the capability of switching between UV-sensing and humidity-sensing devices by shape-reconfiguration via mechanical twisting. It can also serve as an axially foldable and deployable sensing device. The OEM based on the classic Miura-ori pattern realized a biaxially foldable and deployable UV sensing surface that can work at room and extreme temperatures. The simple circular and letter OEMs were able to provide thermal sensing and manipulation functions. The Tachi-Miura OEM was used as an electronic skin attached to the host origami structure to monitor its motion.

The results reported here demonstrated the feasibility of using the designs and principles of engineering origami to advance the state-of-the-art in multifunctional, shape-reconfigurable flexible electronics. This work warrants a few future research and development directions to mature the concept of origami electronic membranes and further explore the applications of origami engineering in the field of flexible electronics. In this work, the OEMs were manually folded and deployed. The first interesting direction in the future is to explore robotic or automatic actuation of the OEMs via motorized actuators or active materials. Second, an interesting application of this concept is for space explorations that pose demanding requirements on the size, weight, and expandable volume. The durability of OEMs in harsh space environments and the methods to protect the OEMs against degradation in space will need to be studied. Furthermore, the scalable manufacturing methods to produce large OEMs at meter scales are also important future research and development topics to study.

## Declaration of competing interest

The authors declare that they have no conflict of interest or financial conflicts to disclose.

# Acknowledgment

This work was partially supported by the Air Force Office of Scientific Research under the award FA9550-22-1-0284. Any opinions, findings, and conclusions or recommendations expressed in this material are those of the author(s) and do not necessarily reflect the views of the United States Air Force. This work was also partially supported by the National Science Foundation (award number 2030579), the Haythornthwaite Foundation Research Initiation Grant, and the startup funds from the University of Illinois at Urbana-Champaign. Li acknowledges the financial support from the Pennsylvania State University.

# References

(1) Huang, Z.; Hao, Y.; Li, Y.; Hu, H.; Wang, C.; Nomoto, A.; Pan, T.; Gu, Y.; Chen, Y.; Zhang, T.; et al. Three-Dimensional Integrated Stretchable Electronics. *Nat. Electron.* **2018**, *1*, 473-480.
(2) Sim, K.; Chen, S.; Li, Z.; Rao, Z.; Liu, J.; Lu, Y.; Jang, S.; Ershad, F.; Chen, J.; Xiao, J.; et al. Three-Dimensional Curvy Electronics Created using Conformal Additive Stamp Printing. *Nat. Electron.* **2019**, *2*, 471-479.
(3) Ko, H. C.; Stoykovich, M. P.; Song, J.; Malyarchuk, V.; Choi, W. M.; Yu, C. J.; Geddes, J. B.; Xiao, J.; Wang, S.; Huang, Y.; et al. A Hemispherical Electronic Eye Camera Based on Compressible Silicon Optoelectronics. *Nature* **2008**, *454*, 748-753.
(4) Ko, H. C.; Shin, G.; Wang, S.; Stoykovich, M. P.; Lee, J. W.; Kim, D. H.; Ha, J. S.; Huang, Y.; Hwang, K. C.; Rogers, J. A. Curvilinear Electronics Formed Using Silicon Membrane Circuit and E;astomeric Transfer Elements. *Small* **2009**, *5*, 2703-2709.
(5) Kim, J.; Kim, M.; Lee, M. S.; Kim, K.; Ji, S.; Kim, Y. T.; Park, J.; Na, K.; Bae, K. H.; Kim, H. K.; et al. Wearable Smart Sensor Systems Integrated on Soft Contact Lenes for Wireless Ocular Diagnostics. *Nat. Commun.* **2017**, *8*, 14997.
(6) Ahn, J. H.; Kim, H. S.; Lee, K. J.; Jeon, S.; Kang, S. J.; Sun, Y.; Nuzzo, R. G.; Rogers, J. A. Heterogeneous Three-Dimensional Electronics by Use of Printed Semiconductor Nanomaterials. *Science* **2006**, *314*, 1754-1757. Javey, A.; Friedman, R. S.; Yan, H.; Lieber, C. M. Layer-by-Layer Assembly of Nanowires for Three-Dimensional, Multifunctional Eectronics. *Nano Lett.* **2007**, *7*, 773-777.
(7) Song, Z.; Ma, T.; Tang, R.; Cheng, Q.; Wang, X.; Krisharaju, D.; Panat, R.; Chan, C. K.; Yu, Y.; Jiang, H. Origami Lithium-Ion Batteries. *Nat. Commun.* **2014**, *5*, 3140.
(8) Xu, S.; Yan, Z.; Jang, K. I.; Huang, W.; Fu, H.; Kim, J.; Wei, Z.; Flavin, M.; McCracken, J.; Wang, R.; et al. Assembly of Micro/Nanomaterials into Complex, Three-Dimensional Architectures by Compressive Buckling. *Science* **2015**, *347*, 154-159.
(9) Miura, K. Method of Packing and Deployment of Large Membranes in Space. *Inst. Space Astronaut. Sci. Rep* **1985**, 168.

(10) Reis, P. M.; Jimémez, F. L.; Marthelot, J. Transforming Architectures Inspired by Origami. *PNAS* **2015**, *6*, 12234-12235.
(11) Rus, D.; Tolley, M. T. Design, Fabrication and Control of Origami Robots. *Nat. Rev. Mater.* **2018**, *3*, 101-112.
(12) Martinez, R. V.; Fish, C. R.; Chen, X.; Whitesides, G. M. Elastomeric Origami: Programmable Paper-Elastomer Composites as Pneumatic Actuators. *Adv. Funct. Mater.* **2012**, *22*, 1376-1384.
(13) Silverberg, J. L.; Evans, A. A.; McLeod, L.; Hayward, R. C.; Hull, T.; Santangeglo, C. D.; Cohen, I. Using Origami Design Principles to Fold Reprogrammable Mechanical Metamaterials. *Science* **2014**, *345*, 647-650.
(14) Kuribayashi, K.; Tsuchiya, K.; You, Z.; Tomus, D.; Umemoto, M.; Ito, T.; Sasaki, M. Self-Deployable Origami Stent Grafts as a Biomedical Application of Ni-Rich TiNi Shape Memory Alloy Foil. *Mater. Sci. Eng.* **2006**, *419*, 131-137.
(15) Lin, C. H.; Tsai, D. S.; Wei, T. C.; Lien, D. H.; Ke, J. J.; Su, C. H.; Sun, J. Y.; Liao, Y. C.; He, J. H. Highly Deformable Origami Paper Photodetector Arrays. *ACS Nano* **2017**, *11*, 10230-10235.
(16) Zirbel, S. A.; Lang, R. J.; Thomson, M. W.; Sigel, D. A.; Walkemeyer, P. E.; Trease, B. P.; Magleby, S. P.; Howell, L. L. Accommodating Thickness in Origami-Based Deployable Arrays1. *Journal of Mechanical Design* **2013**, *135* (11). DOI: 10.1115/1.4025372 (acccessed 5/28/2024).
(17) Berkebile, D. H.; Stevenson, D. L. The Use of "Kapton" Polyimide Film in Aerospace Applications. *SAE Transactions* **1981**, *90*, 3562-3568. (acccessed 2024/05/28/).JSTOR. Arakawa, E. T.; Williams, M. W.; Ashley, J. C.; Painter, L. R. The optical properties of Kapton: Measurement and applications. *Journal of Applied Physics* **1981**, *52* (5), 3579-3582. DOI: 10.1063/1.329140 (acccessed 5/28/2024).
(18) Gouzman, I.; Grossman, E.; Verker, R.; Atar, N.; Bolker, A.; Eliaz, N. Advances in Polyimide-Based Materials for Space Applications. *Advanced Materials* **2019**, *31* (18), 1807738. DOI: https://doi.org/10.1002/adma.201807738.
(19) Jianguo, C.; Xiaowei, D.; Ya, Z.; Jian, F.; Yongming, T. Bistable Behavior of the Cylindrical Origami Structure With Kresling Pattern. *Journal of Mechanical Design* **2015**, *137* (6). DOI: 10.1115/1.4030158 (acccessed 5/28/2024).
(20) Miura, K. Method of Packaging and Deployment of Large Membranes in Space. 1985.
(21) Chen, Y.; Peng, R.; You, Z. Origami of thick panels. *Science* **2015**, *349* (6246), 396-400. DOI: doi:10.1126/science.aab2870.
(22) Miura, K.; Pellegrino, S. *Forms and Concepts for Lightweight Structures*; Cambridge University Press, 2020. DOI: DOI: 10.1017/9781139048569.
(23) Yasuda, H.; Yein, T.; Tachi, T.; Miura, K.; Taya, M. Folding behaviour of Tachi–Miura polyhedron bellows. *Proceedings of the Royal Society A: Mathematical, Physical and Engineering Sciences* **2013**, *469* (2159), 20130351. DOI: doi:10.1098/rspa.2013.0351.
(24) Yao, Y.; Ning, X. Soft electronic skin for self-deployable tape-spring hinges. *Communications Engineering* **2024**, *3* (1), 16. DOI: 10.1038/s44172-024-00163-x.
(25) Li, G.; Yao, Y.; Ashok, N.; Ning, X. Ultra-Flexible Visible-Blind Optoelectronics for Wired and Wireless UV Sensing in Harsh Environments. *Advanced Materials Technologies* **2021**, *6* (9), 2001125. DOI: https://doi.org/10.1002/admt.202001125.

# Supporting Information

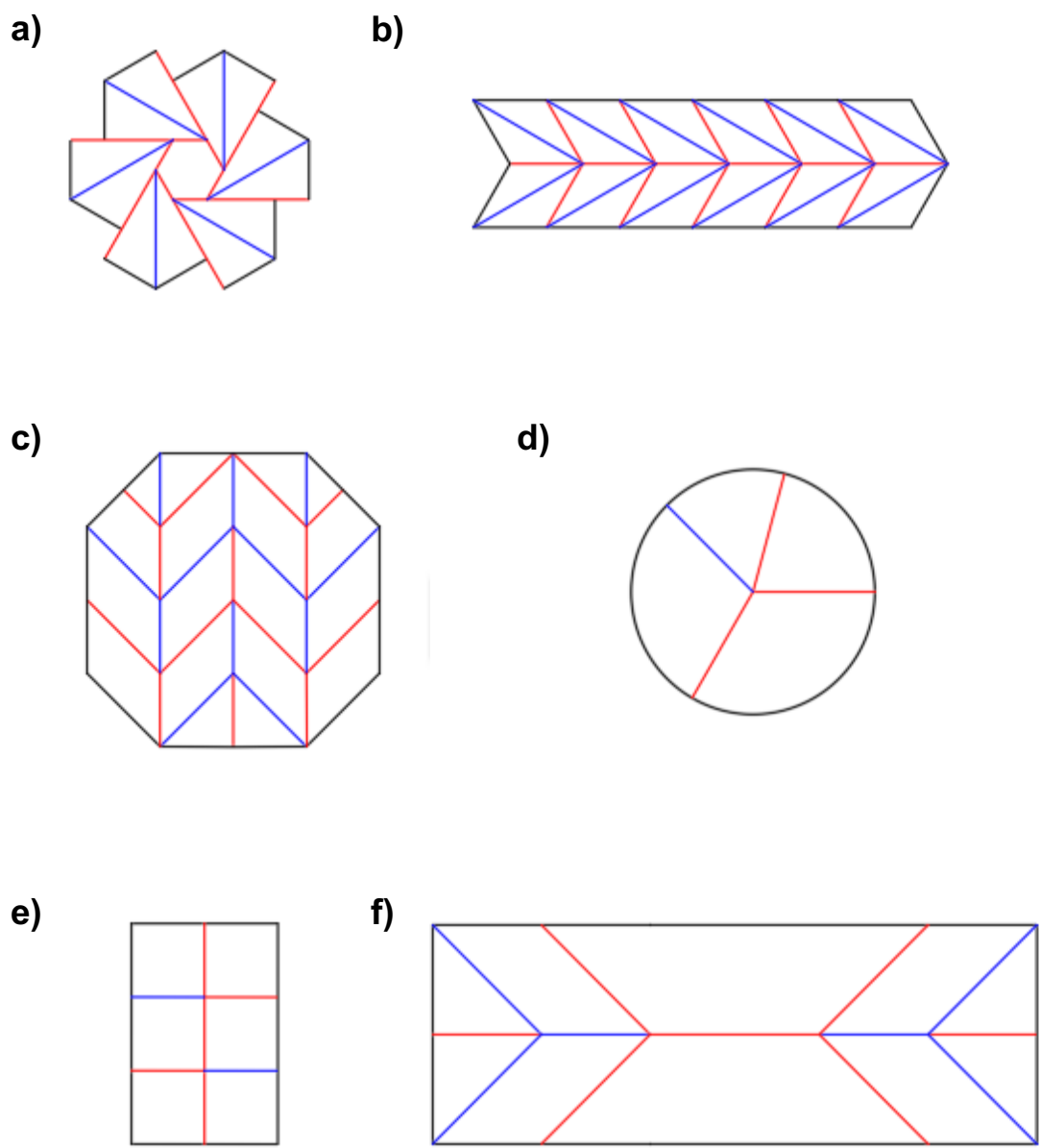

**Figure S1. Origami folding patterns.** a) Modified flasher pattern. b) Kresling pattern. c) Miura-Ori folding pattern. d) Simple membrane folding pattern. e) Letter folding pattern. f) Tachi-Miura folding pattern. (Blue lines: valley folds; Red lines: hill folds.)

| | Excitation Type | Sensor Material | Sensor Type |
|---|---|---|---|
| **Environmental** | Temperature Fluctuation | Copper | Thin-film |
| | Ultraviolet Light Incidence | Zinc Oxide/ Carbon Nanotube | |
| | Humidity Variation | Reduced Graphene Oxide | |
| **Structural** | In-plane Strain | Reduced Graphene Oxide/ Carbon Nanotube | |
| | Mechanical Movement | Microchips | Micro-electro-mechanical Systems |

**Table S1.** Summary of environmental and structural sensing modalities and materials developed and used for origami electronic membranes (OEMs).

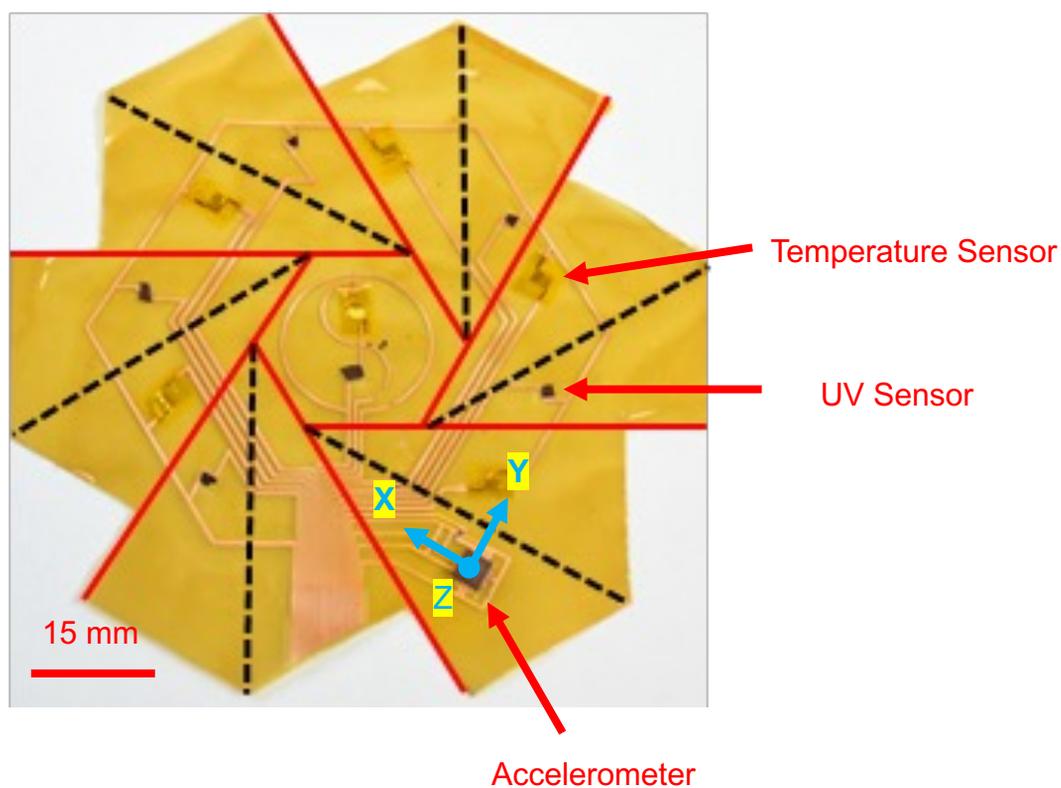

**Figure S2. Optical images of an origami electronic membrane based on a modified flasher pattern in extended state.** The locations of temperature sensors, UV sensor, and the accelerometer are marked. The relative orientation of the accelerometer is marked on the image.

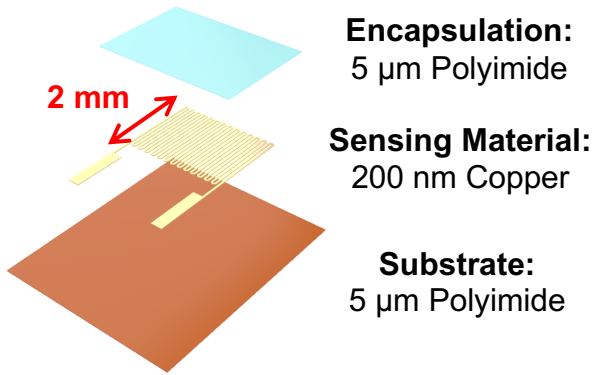 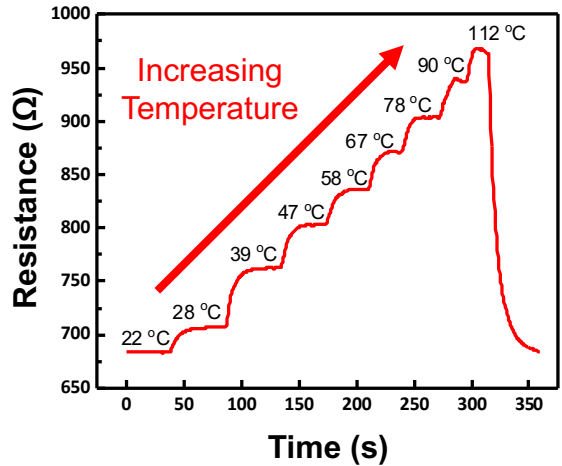

**Figure S3. Design and nominal response of thin-film temperature sensor.**

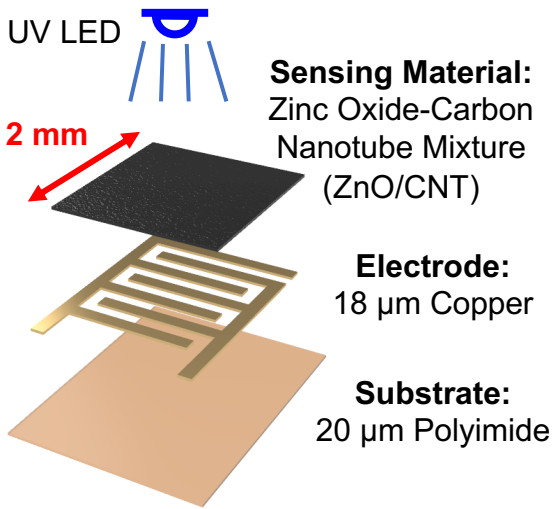 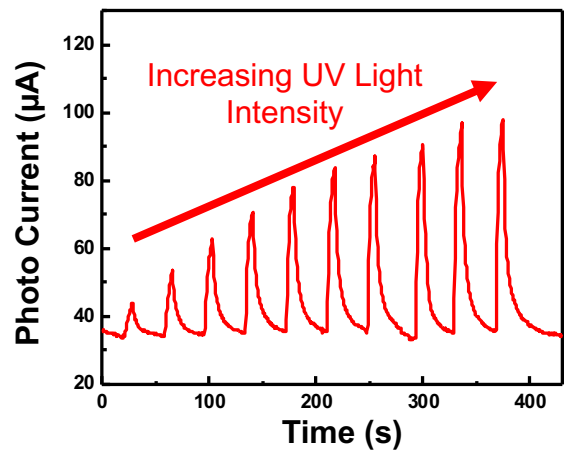

**Figure S4. Design and nominal response of thin-film UV sensor.**

**Thin-Film Humidity Sensor**

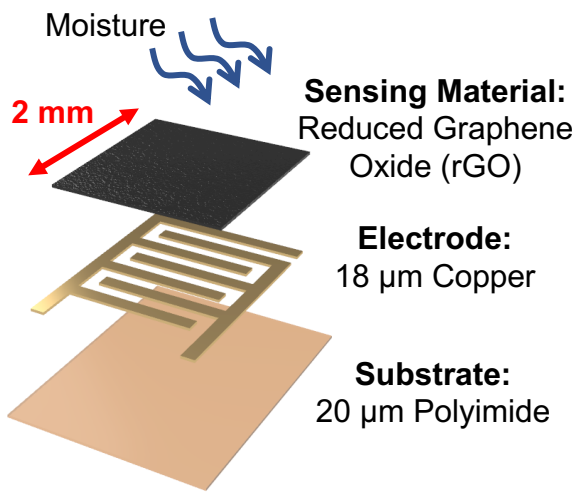

**Humidity Sensor Nominal Response**

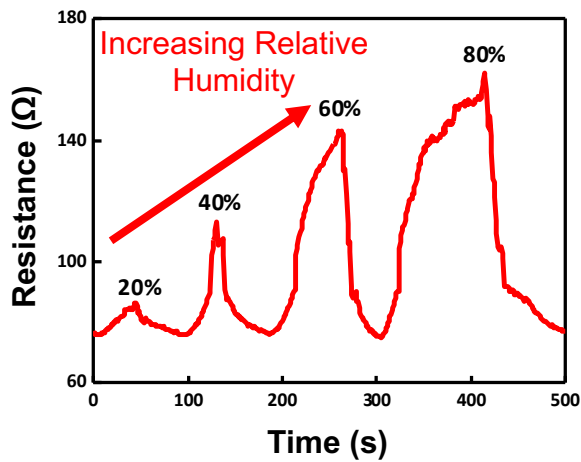

**Figure S5. Design and nominal response of thin-film humidity sensor.**

**Thin-Film Strain Gauge**

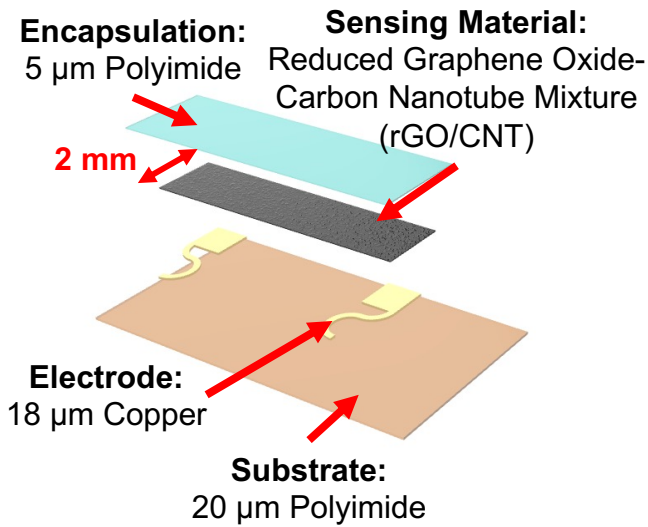

**Strain Gauge Nominal Response**

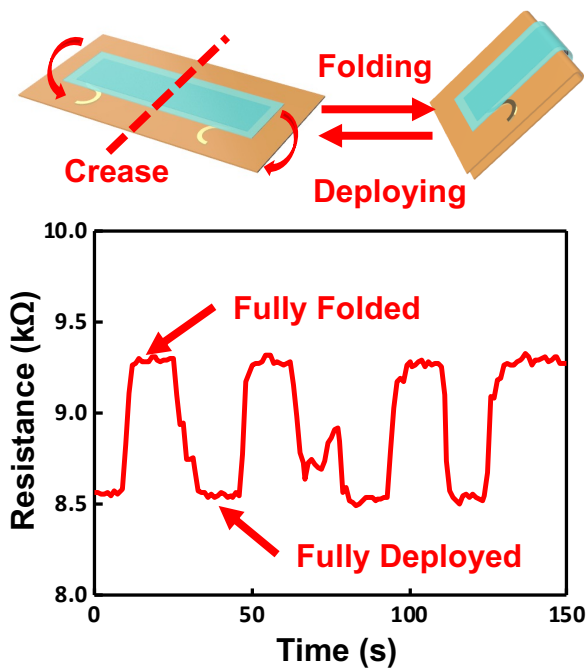

**Figure S6. Design and nominal response of thin-film strain gauge.**